\documentclass[conference]{IEEEtran}
\usepackage[]{amssymb}
\usepackage{graphicx}
\ifCLASSINFOpdf
\else
\fi
\newtheorem{Theorem}{Theorem}
\newtheorem{Proposition}{Proposition}

\newtheorem{Definition}{Definition}
\newtheorem{Example}{Example}

\newtheorem{Remark}{Remark}

\newcommand{\firkant}{\begin{flushright} \vspace{-.7cm} $\Box$ \end{flushright}}

\begin{document}
%
\title{A Note on the Injection Distance}



%
\author{\IEEEauthorblockN{Stanislav Bulygin\IEEEauthorrefmark{1},
Olav Geil\IEEEauthorrefmark{2} and
Diego Ruano\IEEEauthorrefmark{2}}
\IEEEauthorblockA{\IEEEauthorrefmark{1}Center for Advanced Security Research Darmstadt, Technische Universit\"{a}t Darmstadt, Germany\\ Email: Stanislav.Bulygin@cased.de}
\IEEEauthorblockA{\IEEEauthorrefmark{2}Department of Mathematical Sciences, Aalborg University, Denmark\\
Email: \{olav,diego\}@math.aau.dk}
}


\maketitle

\begin{abstract}
K\"{o}tter and Kschischang showed in~\cite{kk} that the network coding counterpart of  Gabidulin codes performs asymptotically optimal with respect to the subspace distance. Recently, Silva and Kschischang introduced in~\cite{sk} the injection distance to give a detailed picture of what happens in noncoherent network coding. We show that the above codes are also asymptotically optimal with respect to this distance.

\end{abstract}


%
\IEEEpeerreviewmaketitle

\section{Introduction}

The concept of error correction in linear network coding was introduced by Cai and Yeung in \cite{cai1}, \cite{cai2} and \cite{cai3}. The scenario considered there is known as coherent network coding meaning that the network topology as well as the linear network code are assumed to be known both to the sender and to the receivers.

Noncoherent network coding was considered by K\"{o}tter and
Kschischang in~\cite{kk} for the first time. Here, neither the
sender nor the receivers are assumed to know the topology or the
linear network code. To mimic the communication situation K\"{o}tter
and Kschischang coined the concept of an operator channel which
takes as input and output subsets of some fixed ambient vector space
$W$. The set of vector subspaces is denoted by ${\mathcal{P}}(W)$.
The game of error correction now  is to identify the set of messages
with a collection of subspaces $C \subseteq {\mathcal{P}}(W)$ called
a subspace code. If $C$ has been chosen cleverly it will, under
certain assumptions, be possible to recover the message at the
receiving end by performing some decoding algorithm. As part of
their description they introduced the subspace distance in
${\mathcal{P}}(W)$. Using this distance then a minimum distance of
$C$ is obtained.  Let $t$ be the number of errors and $\rho$ be the
number of erasures occurred during the transmission over the channel (we will not formally
define here what these concepts mean). The original message can be
recovered if $2(t+\rho)$ is less than the minimum distance of the
subspace code under consideration. However, the converse does not
necessarily hold. K\"{o}tter and Kschischang adapted the rank-metric
code construction by Gabidulin to work in the above setting and gave
an efficient decoding algorithm for them. They presented a Singleton
bound and demonstrated that the adapted Gabidulin codes attain it
asymptotically.

In~\cite{sk} Silva and Kschischang considered a slightly different
model of non-coherent network coding. In particular they coined a
new distance, namely the injection distance. Their interpretation of
the number $t$ of errors and the number $\rho$ of erasures also
differs from the one in~\cite{kk}. The advantage of the model
in~\cite{sk} is that it allows not only a sufficient, but also a
necessary condition for decoding to be possible. We now describe the
model in detail. As above each message is identified with a codeword
in $C \subseteq {\mathcal{P}}(W)$. The sender injects  a possibly
overcomplete basis for this subspace into the network. The nodes
then forward a linear combination of the incoming vectors on each
outgoing edge and possibly add an error vector. Let $X$ be an $n
\times m$  matrix which rows are the source packets and for a
specific receiver denote by $Y$ the $N \times m$ matrix which rows
are the received packets. Let the number of error vectors be $t$ and
denote by $Z$ the $t \times m$ matrix which rows are the errors.
With respect to the specific receiver let $A$ be the $N \times n$
transfer matrix for the linear code (the error free part) and let
$D$ be the $N \times t$ transfer matrix for the errors. This gives
us the model
$$Y=AX+DZ.$$
The transfer matrices $A$ and $D$ are unknown to the receiver and are chosen by the adversary while respecting the constraint ${\mbox{rank}}( A) \geq n-\rho$. Here, $\rho$ is a parameter called the rank deficiency of $A$, known to all the participants. \\
Knowing $Y$ the decoding rule to use is
$$\hat{X}={\mbox{argmin}}_{X \in C} \Delta_{\rho}(X,Y)$$
where
\begin{eqnarray}
&&\Delta_{\rho}(X,Y)\nonumber \\
&=&\min \{ r \mid
A \in {\mathbf{F}}_q^{N \times n}, r \in {\mathbf{N}}, D \in
{\mathbf{F}}_q^{N \times r},\nonumber \\
&&{\mbox{ \ \ \ \ }}  Z \in {\mathbf{F}}_q^{r \times m},~
Y=AX+DZ, ~
{\mbox{rank}}(A) \geq n-\rho \}\nonumber \\
&=&\max \{ \dim (X)-\rho,\dim Y \} - \dim (X \cap Y). \nonumber
\end{eqnarray}
Here, the last equality corresponds to \cite[Theorem 16]{sk}.
The ability of a subspace code $C$ to support the above decoding algorithm is described by the following parameter.
\begin{Definition}
The injection distance between spaces $U,V \in {\mathcal{P}}(W)$ is defined as follows
$$d_I(U,V)=\dim (U+V)-\min \{ \dim (U), \dim(V) \}.$$
This in an obvious way translates into a minimum distance $d(C)$ for
any subspace code $C\subseteq {\mathcal{P}}(W)$.
\end{Definition}
\begin{Theorem}
Assume there is a bijective map between the set of messages and the
subspace code $C$. The code is guaranteed to correct $t$ packet errors, under rank deficiency $\rho$, if and only if $d_I(C)> 2t +\rho$.
\end{Theorem}
The injection distance relates to the subspace distance $d_S$ from~\cite{kk} as follows
$$d_I(U,V)=\frac{1}{2} d_S(U,V)+\frac{1}{2} | \dim (U)- \dim (V) |.$$
Hence, except for a factor $\frac{1}{2}$ the two distances are the same if $\dim
(U)= \dim (V)$. A subspace code $C$ is called equidimensional if all
of its codewords have the same fixed dimension. It is clear that except for
a factor $\frac{1}{2}$ the minimum distance of an equidimensional
subspace code is the same no matter which of the metrics $d_I$ or $d_S$ is used.

\section{Asymptotic results}

A subspace code $C \subset \mathcal{P}(W)$ with $W$ an $N$-dimensional
vector space over $\mathbb{F}_q$ (the finite field with $q$ elements), with size $| C |$, maximum dimension
of a codeword $l=\max_{x\in C} \dim (X)$ and minimum injection
distance $D=d_I (C)$ is said to be of type [$N,l,\log_q | C |,
D$]. The parameter
$$
R=\frac{\log_q(|C|)}{Nl}
$$
is called the rate of the code (see \cite[Definition 2]{kk}). This
parameter clearly serves as a measure of the efficiency of
communication in a model where every block of information consists of
$l$ vectors of size $N$ that are injected into the system. In other
words, in a situation where we inject a possible
overcomplete basis of $l$ vectors into the system, the dimension of
the codeword is unimportant. Following the ideas of~\cite{kk} we will give a
Singleton type upper bound on $|C|$ in terms of $l, D$ and $N$. This
will then give us an upper bound for $R$ which we finally show to be
reached asymptotically by the network coding counterparts of Gabidulin codes.\\

We consider the definition of puncturing from \cite{kk}.

\begin{Definition}
Let $C \subset \mathcal{P} (W)$, with $\dim (W) =N$ and let $W'$ be a subspace of $W$ of dimension $N-1$. A punctured code $C'$ is constructed from $C$ by replacing $V \in C$ by $V' = \mathcal{H}_{\dim(V)-1} (V \cap W')$. That is,

\begin{itemize}
\item $V' = V \cap W'$, if $V \cap W'$ has dimension $\dim (V) -1$.
\item $V'$ a random subspace of dimension $\dim (V) -1$, otherwise.
\end{itemize}We remark that the definition of $C'$ is not unique.
\end{Definition}

This definition allows us to extend \cite[Theorem 8]{kk} for the injection distance.

\begin{Proposition}\label{pro:pun}
  Let $C$ be a $[N,l,\log_q | C | , D]$ code with $d_I (C)= D > 1$. Then a punctured code $C'$ is of type $[N-1,l-1,\log_q | C | , D']$, with  $D'\ge D-1$.
\end{Proposition}
\noindent   {\it{Proof:}}\ \ It is clear that $\dim W' = N-1$ and the maximum dimension of the codewords is $l-1$.

Let $U,V \in C$ with $U \neq V$, $l_1 = \dim (U)$ and $l_2 = \dim (V)$. Let $U' = \mathcal{H}_{l_1 -1} (U)$ and $V' = \mathcal{H}_{l_2 -1} (V)$.
One has that $\dim(U\cap V) \le \max \{l_1, l_2\} -D$ and therefore $\dim(U'\cap V') \le \dim(U\cap V) \le \max \{l_1, l_2\} -D$ since $U' \subset
U$ and $ V' \subset V$. Hence,
\begin{eqnarray*}
  d(U',V') & = & \max \{l_1 -1, l_2 -1\} - \dim (U' \cap V')\\
  &=& \max \{l_1 , l_2 \} -1 - \dim (U' \cap V')\\
  &\ge & D -1
\end{eqnarray*}
Since $D > 1$, we have as many codewords in $C'$ as in $C$.\firkant

For nonnegative integers $l, n$ with $l \le n$, the $q$-ary Gaussian coefficient is
$$
 \left [ {N \atop l}  \right ]_q = \prod_{i=0}^{l-1} \frac{q^{N-i} -1}{q^{l-i}-1}
$$and for $l=0$ is defined to be $1$. The number of vector subspaces
of dimension $l$ of an $N$-dimensional vector space is given by $\left
  [ {N \atop l}  \right ]_q$. We may establish a Singleton type bound
for codes with $l \le N/2$ and the injection distance. This result
extends \cite[Theorem 9]{kk} where a Singleton bound
is established for the subspace distance and equidimensional codes.

\begin{Theorem}\label{th:sbi}
Let $C$ be a   [$N,l,\log_q| C |,D $] code, with $l \le N/2$. Then, $$| C | \le  1+ (l -D+1) \left [ {N-D+1 \atop N-l}  \right ]_q.$$
\end{Theorem}
\noindent {\it{Proof:}} \ \  We can puncture $D-1$ times the code $C$ to obtain a code $C'$ of
type [$N-(D-1),l-(D-1),\log_q| C |,D' $] , with $D'\ge 1$, by Proposition \ref{pro:pun} (if
$D=1$ we do not puncture it and $C=C'$). One has that $C' \subset
W'$, with $\dim (W') = N - D +1$. We bound the number of subspaces
of $W'$ with dimension lower than or equal to $l-D+1$:
\begin{eqnarray*}| \mathcal{P}(W',\le l-D+1) | & =  &
\sum_{i=0}^{l-D+1} \left [ {N-D+1 \atop i}  \right ]_q \\ & \le & 1
+ (l-D+1) \left [ N-D+1 \atop l - D +1  \right ]_q \\ & = & 1+ (l
-D+1) \left [ {N-D+1 \atop N-l}  \right ]_q,\\ \end{eqnarray*}since
$\left [ {N \atop l}  \right ]_q = \left [ {N \atop n-l}  \right
]_q$.

\firkant

For a code with $l > N/2$ a Singleton bound would be a trivial bound
because, for $N$ fixed, $\left [ {N \atop l}  \right ]_q$ is a
symmetric function on $l$ which has a maximum at $l=N/2$, for $N$
even, or two maximums at $l=(N-1)/2,(N+1)/2$, for $N$ odd.

\begin{Remark}
In \cite[Theorem 9]{kk}, for equidimensional codes and the subspace
distance, a Singleton bound is obtained by considering a punctured
code and bounding the number of vector subspaces. The same argument
is considered for the dual code, since $d_S (U,V) = d_S
(U^\bot,V^\bot)$. The bound is the minimum of these two values.

Although for $U, V \subset \mathcal{P}(W)$, $d_I (U,V) = d_I (U^\bot
, V^\bot)$, one cannot consider the dual of $C$ in the proof of the
Singleton bound. Namely, let $C$ be a code with dimension $l \le
N/2$ and let $l_1< \cdots < l_s=l$ the dimension of the words of
$C$. One has that the words of $C^\bot$ have dimension $N-l_s,
\ldots , N-l_1$, and the words of its punctured code have dimension
$N-l_s-D+1, \ldots , N-l_1-D+1$.

Let $C$ be a code with $l=N/2 -1$. Then the dimension of the
smallest codeword of $C^\bot$ is $l_1^\bot = N/2 +1$. After
puncturing it $D-1$ times, one has that the dimension of the
smallest codeword is $N/2 -D +2$. Then,$$| C'^\bot | \le
\sum_{i=0}^{N/2 -1} \left [ N-(D-1) \atop N-i - (D-1)  \right
]_q,$$however, it is not clear which of the Gaussian coefficients it
the largest one. For instance, the largest Gaussian coefficient is
$\left [ N-(D-1) \atop N- (N/2 -1) -(D-1) \right ]$ (for $(N/2 - D
+2$  even) if and only if $N/2 -(D-1)/2 < N/2 + 2 - D$, that is, $D
<5$.
\end{Remark}

We consider the codes from \cite[Section V-B]{kk}, these codes are
the translation of the Gabidulin rank-matrix code construction to
subspace codes, these codes nearly achieve the Singleton bound for
the subspace distance. Moreover, they verify the assumption $l \le
N/2$ and we claim that they also nearly achieve the Singleton-Bound
for the injection distance (Theorem \ref{th:sbi}).

Gabidulin codes are equidimensional which allows us to calculate
their injection distance as follows $d_I (C)= d_S(C)/2=l-k+1$. Here,
the last part is from~\cite{kk}. They are of type
$[l+m,l,mk,l-k+1]$, with $l \le m$ and $l \ge k$. We have that $l
\le N/2$, since  $l \le N/2$ if and only if $l \le (l+m)/2$, that
is, $l \le m$.

We consider the bound from theorem \ref{th:sbi},
\begin{eqnarray}
| C | & \le & 1 + (l-(l-k+1)+1) \left [ N-((l-k+1)-1) \atop N-l
\right ]_q\nonumber \\
& =& 1 + k\left [ l+m -l +k \atop l + m -l \right ]_q\nonumber \\
& = & 1 + k \left [ m+k \atop m \right ]_q \label{eq3gange} \\
& < & 1 +  4 kq^{mk}. \label{eq4gange}
\end{eqnarray}
The last inequality follows from $1 < q^{l(N-l)}\left[ N \atop l  \right ]_q < 4$ (see \cite[Lemma 4]{kk}).

Therefore, a code achieving the bound in Theorem~\ref{th:sbi} cannot
have more than $4k$ times as many codewords as a Gabidulin code. Consider now
the rate corresponding to~(\ref{eq4gange})
$$R=\log_q(|C|)/Nl=\frac{\log_q(\frac{1}{q^{mq}}+4k)+km }{Nl}.$$
As by construction $k \leq l$ this tends to $\frac{km}{Nl}$ as $N$ goes
to infinity. In other words the network coding counterpart of
Gabidulin codes  asymptotically has the maximal rate. The next example
illustrates that already for small $N$ the Gabidulin codes perform quite well.
\begin{Example}
In this Example we consider codes of the Gabidulin type over the
field ${\mathbb{F}}_{16}$, the finite field with $16$ elements. We
consider a sequence of values
$$\{[m_4,l_4,k_4],[m_5,l_5,k_5], \ldots ,[m_{30},l_{30},k_{30}]\}$$
with $m_i=i$, $l_i=\lfloor \frac{3m_i}{5}\rfloor$ and $k_i=\lfloor
\frac{m_i}{2}\rfloor$ for $i=4, \ldots , 30$.
The corresponding codes  have length $N$ equal to
$$\{6,8,9,\ldots ,48\}.$$
In Figure~\ref{nogetandet}, the rates of the Gabidulin type codes are plotted with $\diamond$'s . The $+$'s correspond to the upper
bound~(\ref{eq3gange}) and the $\circ$'s correspond to the upper bound~(\ref{eq4gange}).

\begin{figure}[h]
\begin{center}
\includegraphics[width=8cm]{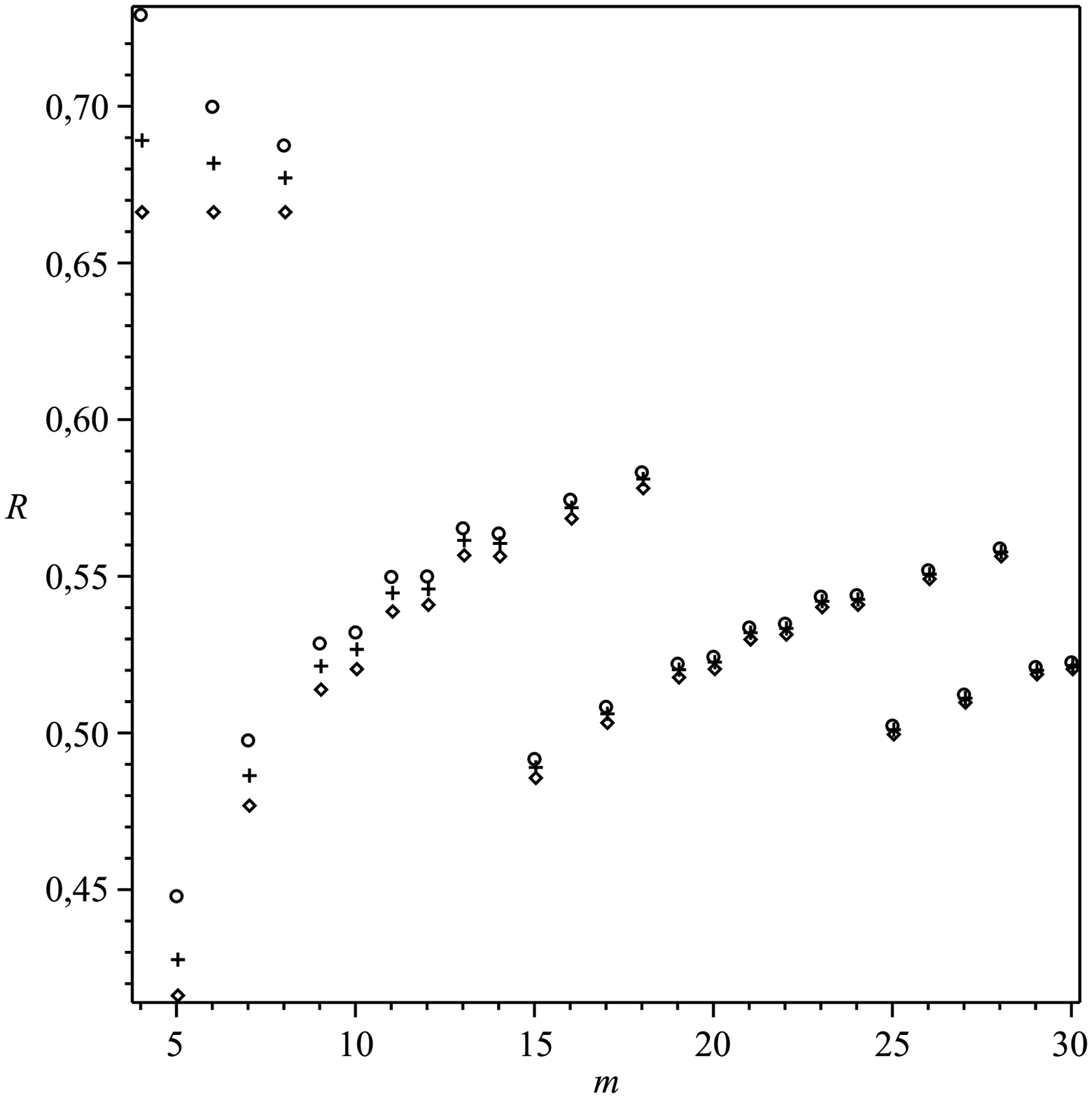}
\end{center}
\caption{}
\label{nogetandet}
\end{figure}
\end{Example}


\begin{thebibliography}{99}


\bibitem{kk} R.~K\"{o}tter and F.~R.~Kschischang, ``Coding for Errors and Erasures in Random Network Coding,'' {\textit{IEEE Trans.~Inform.~Theory}}, {\textbf{54(8)}}, 2008, pp.~3579-3591.


\bibitem{sk} D.~Silva and F.~R.~Kschischang, ``On Metrics for Error Correction in Network Coding,'' {\textit{ArXiv}}:0805.3824v4[cs.IT], 2009, To appear in {\textit{IEEE Trans.~Inform.~Theory,}} 28 pages.

\bibitem{cai1} N.\ Cai and R.\ W.\ Yeung, ``Network coding and error correction,'' in {\textit{Proc.\ 2002, IEEE Inform.\ Theory Workshop,}} Oct.\ 20-25, 2002, pp.\ 119-122.


\bibitem{cai2} R.\ W.\ Yeung and N.\ Cai, ``Network error correction, part I: Basic concepts and upper bounds,'' {\textit{Commun.\ Inform.\ Syst.,}} {\textbf{6}}, No.\ 1, 2006, pp.\ 19-36.


\bibitem{cai3} N.\ Cai and R.\ W.\ Yeung, ``Network error correction, part II: Lower bounds,'' {\textit{Commun.\ Inform.\ Syst.,}} {\textbf{6}}, No.\ 1, 2006, pp.\ 37-54.




\end{thebibliography}
\end{document}